\documentclass[11pt]{article}

\usepackage[preprint]{acl}

\usepackage{times}
\usepackage{latexsym}
\usepackage[T1]{fontenc}
\usepackage[utf8]{inputenc}
\usepackage{microtype}
\usepackage{inconsolata}
\usepackage{graphicx}
\usepackage{booktabs}
\usepackage{multirow}
\usepackage{array}
\usepackage{amsmath}
\usepackage{amssymb}
\usepackage{xcolor}
\usepackage{url}
\usepackage{enumitem}
\usepackage{placeins}
\usepackage{listings}
\usepackage{tikz}
\usetikzlibrary{shapes,arrows.meta,positioning,fit,backgrounds}

\graphicspath{{figures/}}
\title{Glite ARF: Verifier-Driven Research with\\
Parallel LLM Coding Agents}

\author{
  Vassili Philippov$^{1}$, Pavel Katunin$^{1}$,
  Dmitry Andreev$^{1}$, Igor Ostanin$^{1}$,
  Anton Nikolaev$^{2}$ \\
  $^{1}$Glite, \texttt{\{vassili,pavel,dmitrii,igor\}@glite.ai} \\
  $^{2}$School of Biosciences, The University of Sheffield, Sheffield, UK \\
  \texttt{a.nikolaev@sheffield.ac.uk}
}

\hypersetup{
  pdftitle={Glite ARF: Verifier-Driven Research with Parallel LLM Coding Agents},
  pdfauthor={Vassili Philippov, Pavel Katunin, Dmitry Andreev, Igor Ostanin, Anton Nikolaev}
}

\begin{document}
\maketitle

\begin{abstract}
LLM coding agents make it tempting to automate empirical research by
delegating experiments to them directly, but naive delegation does not
scale to large projects: low-rate instruction lapses compound into
broken, irreproducible artefacts. To address this problem, we present
\textbf{Glite ARF}, an open-source Python framework for running many
LLM coding agents in parallel on a research repository without
sacrificing reproducibility or auditability. The framework defines a
\textbf{three-role stack}: a
human researcher chooses which hypotheses to test, coding agents
(Claude Code, Codex CLI) implement individual tasks under a fixed
structure, and deterministic Python \emph{verifier} scripts enforce task isolation, immutability of completed
work, a corrections overlay, and a materialised project overview. We
call this
\textbf{verifier-driven research}: the rules of the research
process live in code that fails loudly when violated, not in prose
that agents are merely asked to follow. Using Glite ARF, we developed
our submission to the BEA~2026 vocabulary-difficulty shared task,
placing first in the closed track and second in the open track on all
three target languages (Spanish, German, Mandarin) and reducing the
official baseline RMSE by 29.9\% (closed) and 35.9\% (open). The
campaign comprised 273 tracked tasks (146 experiment runs) across 129
feature sets, run by up to twelve parallel agents orchestrated from a
single laptop --- with some model training on rented A100s --- at
${\sim}\$450$ in LLM API spend (${\sim}\$498$ total third-party cost),
and structured per-fold provenance let us catch and strip four
target-leaking feature sets, correcting an implausible 0.609 RMSE to
0.802. Across three campaigns in three domains, the framework's
structural machinery adds only ${\sim}1\%$ of wall-clock time.
Framework and a public demo project accompany this paper.
\end{abstract}

\section{Introduction}
\label{sec:introduction}

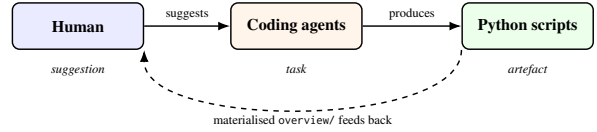
\begin{figure}[t]
\centering
\resizebox{\columnwidth}{!}{%
\begin{tikzpicture}[
  font=\small,
  role/.style={
    draw, rounded corners, thick,
    minimum width=2.55cm, minimum height=0.9cm,
    align=center
  },
  unit/.style={font=\itshape\scriptsize, align=center, fill=white, inner sep=1pt},
  arr/.style={-Latex, thick},
  edge label/.style={font=\scriptsize, fill=white, inner sep=1pt}
]
  \node[role, fill=blue!8]   (human)  at (0,0)   {\textbf{Human}};
  \node[role, fill=orange!8] (agent)  at (4.25,0) {\textbf{Coding agents}};
  \node[role, fill=green!8]  (script) at (8.75,0) {\textbf{Python scripts}};
  \node[unit, below=7pt of human]  {suggestion};
  \node[unit, below=7pt of agent]  {task};
  \node[unit, below=7pt of script] {artefact};
  \draw[arr] (human.east)  -- node[above=2pt, edge label]{suggests} (agent.west);
  \draw[arr] (agent.east)  -- node[above=2pt, edge label]{produces} (script.west);
  \draw[arr, dashed]
    (script.south west) .. controls +(0,-1.55) and +(0,-1.55) ..
    node[below=2pt, edge label, pos=0.5]
      {materialised \texttt{overview/} feeds back}
    (human.south east);
\end{tikzpicture}}
\caption{Glite ARF's three-role stack. The human researcher writes
suggestions; coding agents execute tasks; deterministic Python scripts
verify artefacts and materialise the canonical view that the human
reads back. Each role works at a different unit of granularity.}
\label{fig:three-role-stack}
\end{figure}

LLM coding agents make it natural to imagine research as many
delegated experiments running in parallel. Naive automation, however,
does not scale to large projects: agents follow most instructions, but
the few they skip compound into fabricated citations, contaminated
splits, stale summaries, out-of-scope edits, and irreproducible
provenance. Glite ARF addresses this failure mode by making the
research process itself executable. A human researcher proposes
hypotheses, agents implement isolated tasks, and deterministic Python
\emph{verifiers} enforce task structure, immutability, corrections,
and a materialised project overview (Figure~\ref{fig:three-role-stack}).

We used Glite ARF to develop our submission to the BEA~2026
vocabulary-difficulty shared task \citep{skidmore2025kvl}, detailed in
our case study (\S\ref{sec:case-study}). The resulting system placed first
in the closed track and second in the open track on all three target
languages; we developed it across 273 tracked tasks (146 experiment
runs) inside the Python framework released with this paper. The campaign
covered 129 feature sets at approximately \$450 in LLM API spend for
feature engineering (\$498 total third-party cost, including rented
compute). We have
since run the framework across three campaigns in three domains; its
structural machinery adds only ${\sim}1\%$ of wall-clock time
(\S\ref{sec:in-use}).

The failure mode we target is visible once campaigns involve hundreds
of runs. Independent evaluation of fully autonomous research systems
has shown how the failure rate compounds:
\citet{beel2025evaluating} found that 42\% of experiments
generated by The AI Scientist v1 \citep{lu2024sakana} failed due
to coding errors. Our own audit of the BEA~2026 campaign repository
catalogues thirteen incidents of similar character, ranging from
train-data corruption affecting 38 feature sets to four feature sets
that leaked the target variable into LLM prompts
(Appendix~\ref{app:failure-modes}).

The natural response is to write better prompts. We tried. Across
thousands of agent invocations the failure rate of prompt-encoded
rules does not go to zero --- it stays in the low single digits,
which compounds to dozens of broken artefacts across a multi-week
campaign. We argue that rules in agentic research must live in
deterministic code: scripts that refuse to advance a step until an
artefact conforms to a versioned specification, scripts that detect
when files outside an agent's task folder have been modified, scripts
that materialise the canonical view of cumulative results so no agent
ever needs to write a manual summary. The companion BEA~2026 system
paper describes the same mechanism in its framework-overview section
\citep{glite2026bea}.

Glite ARF defines a \textbf{three-role stack} for autonomous
research (Figure~\ref{fig:three-role-stack}): the human researcher
chooses which hypotheses to test, coding agents implement individual
tasks under fixed structure, and deterministic Python scripts
enforce task isolation, immutability, a corrections overlay, and a
materialised project overview. We call this
\textbf{verifier-driven research}, by analogy with test-driven
development: the rules of the research process live in scripts that
fail loudly when violated, not in prose that agents are merely asked
to follow. Our contribution is the framework, used at scale across
three domains, with a refereed external shared task as one anchor and
measured campaign evidence (\S\ref{sec:in-use}) as another.

Section~\ref{sec:related-work} positions Glite ARF in the
autoresearch and AI-for-science landscape.
Section~\ref{sec:system} describes the framework's seven structural
principles and the verifier layer that enforces them.
Section~\ref{sec:case-study} reports our BEA~2026 case study,
including an example where structured per-fold provenance let us catch
and strip target leakage. Section~\ref{sec:in-use} reports measured
evidence from running the framework across three campaigns.
Section~\ref{sec:lessons} discusses lessons and limits --- most
importantly, we are explicit that role 1 (hypothesis selection)
remains human-driven by design.
Section~\ref{sec:availability} covers availability.

\section{Related Work}
\label{sec:related-work}

\paragraph{Multi-agent orchestration.}
AutoGen \citep{wu2023autogen}, MetaGPT \citep{hong2023metagpt},
CAMEL \citep{li2023camel}, smolagents \citep{huggingface2024smolagents},
and CrewAI \citep{crewai} provide intra-task agent coordination and
may run inside an ARF task; ARF is orthogonal, structuring the
\emph{campaign}-level lifecycle that wraps any of them.

\paragraph{Coding agents.}
Coding agents like Claude Code \citep{anthropic2025claudecode},
OpenAI Codex CLI \citep{openai2025codexcli},
Aider \citep{gauthier2023aider},
OpenHands \citep{wang2024openhands}, and
SWE-agent \citep{yang2024sweagent} execute the code-writing inside an
ARF task. SWE-bench \citep{jimenez2023swebench} is the closest
adjacent benchmark, evaluating single-issue resolution rather than
research-campaign integrity.

\paragraph{Autoresearch (practitioner side).}
A practitioner community has formed around Karpathy's autoresearch
pattern \citep{karpathy2024autoresearch} --- short, single-metric
optimisation loops applied to one training script. Recent work makes
this approach more rigorous: CORAL \citep{qu2026coral} introduces
multi-agent evolution with isolated worktrees and shared persistent
memory; EvoSkill \citep{alzubi2026evoskill} discovers reusable agent
skills through failure-driven evolution with each agent program
represented as a git branch; autoreason
\citep{nous2026autoreason} addresses iterative document refinement by
treating ``do nothing'' as a first-class option.

\paragraph{Autoresearch (academic side).}
On the academic side, AI-Researcher \citep{tang2025airesearcher} and
Agent Laboratory \citep{schmidgall2025agentlab} orchestrate full
literature-to-manuscript pipelines; The AI Scientist v1 and v2
\citep{lu2024sakana,lu2025sakanav2} generate workshop-level papers
end to end; ADAS \citep{hu2025adas} automates the design of agentic
systems themselves. Independent evaluation has shown that fully
autonomous systems still produce substantial structural errors:
\citet{beel2025evaluating} found that 42\% of AI-Scientist-v1
experiments failed due to coding errors. For a survey of the field see
\citet{zheng2025survey}.

\paragraph{Evaluation regime.}
Existing autoresearch systems are typically evaluated against
benchmarks constructed by the system's own team or close
collaborators: MLE-Bench \citep{chan2024mlebench},
MLAgentBench \citep{huang2024mlagentbench},
MLR-Bench \citep{chen2025mlrbench},
AgentBench \citep{liu2024agentbench}, and Scientist-Bench inside
AI-Researcher itself. Glite ARF's empirical anchor is different in
kind: a refereed external shared task (BEA~2026) whose test labels,
metric, and leaderboard adjudication are all outside the authors'
control, complemented by measured evidence from three author-run
campaigns (\S\ref{sec:in-use}).

\paragraph{Workflow, provenance, and experiment management.}
Many of ARF's mechanisms adapt established software- and
data-engineering patterns rather than inventing them. Experiment
trackers (MLflow, Weights \& Biases, Sacred, ClearML, Aim),
data and model versioning (DVC, DataLad), workflow engines
(Snakemake, Nextflow, Kedro), and reproducibility tooling (ReproZip)
already provide provenance, lineage, and pipeline structure;
continuous integration, policy-as-code, and append-only or
event-sourced logs provide gating and immutable history. ARF's
contribution is not any single one of these ideas but their
opinionated, repository-native combination for \emph{parallel
LLM-agent research campaigns}, in which the actor producing artefacts
is a non-deterministic agent rather than a person.
Table~\ref{tab:capabilities} positions ARF against representative
systems along the dimensions this setting stresses.

\begin{table}[t]
\centering
\small
\setlength{\tabcolsep}{4pt}
\begin{tabular}{@{}lcccc@{}}
\toprule
Capability & \rotatebox{90}{Trackers\,} & \rotatebox{90}{DVC/CI\,}
 & \rotatebox{90}{Workflows} & \rotatebox{90}{\textbf{ARF}} \\
\midrule
Task-level worktree isolation        & --- & $\sim$ & ---    & \checkmark \\
Machine-checked artefact contracts   & --- & $\sim$ & $\sim$ & \checkmark \\
Immutable corrections overlay        & --- & ---    & ---    & \checkmark \\
Agent command/transcript logging     & --- & ---    & ---    & \checkmark \\
Materialised human-facing overview   & $\sim$ & --- & ---    & \checkmark \\
Model / agent independence           & \checkmark & \checkmark & \checkmark & \checkmark \\
\bottomrule
\end{tabular}
\caption{ARF versus representative experiment trackers
(MLflow, W\&B), versioning/CI (DVC, GitHub Actions), and workflow
engines (Snakemake). \checkmark\ = provided; $\sim$ = partial;
--- = not provided.}
\label{tab:capabilities}
\end{table}

\paragraph{Position.}
Glite ARF is therefore neither an orchestrator nor an agent nor a
benchmark, but a structural envelope that wraps existing coding agents
and makes a multi-week, multi-agent research campaign auditable.

\section{System}
\label{sec:system}

\subsection{Architecture and seven principles}
\label{sec:system-architecture}

Glite ARF defines three roles. The human researcher works at the
level of \emph{suggestions}: hypotheses to test, datasets to try,
libraries to evaluate. Coding agents (Claude Code, Codex CLI) work at
the level of \emph{tasks}: each task is a folder under
\path{tasks/tNNNN_slug/}, a git branch, and a pull request; large
artefacts such as datasets and model checkpoints live in that folder
under Git LFS by default, though a project can substitute external
object storage (e.g.\ Amazon~S3).
Deterministic Python scripts work at the level of \emph{artefacts}:
every file an agent writes conforms to a versioned specification
(\path{arf/specifications/},
\path{meta/asset_types/<kind>/specification.md}), enforced by a
\emph{verifier} before it is merged; ARF's internal name for these
scripts is \emph{verificator}. ARF
is semi-autonomous by design: role 1 (hypothesis selection) remains
human-driven because frontier models in mid-2026 do not reliably
perform research-direction selection at the timescale of a multi-week
campaign. We return to this design choice in \S\ref{sec:lessons}.

The framework's design crystallised over four research projects in
response to recurring failure modes catalogued in
Appendix~\ref{app:failure-modes}. Seven structural principles emerged.

\paragraph{1. Task isolation.}
Motivated by an incident in which a single agent step intended to
append 2{,}244 test rows instead recomputed train, dev, and test
features across 38 feature sets, corrupting 20{,}304 historical
training rows (Appendix~\ref{app:failure-modes}, F02). Each task lives
in its own folder and git worktree, and may modify only that folder
plus a small allow-list of shared configuration files
(\texttt{pyproject.toml}, \texttt{uv.lock}, \texttt{ruff.toml},
\texttt{mypy.ini}, \texttt{.gitignore}, \texttt{.gitattributes}). A
pre-merge verifier (\texttt{PM-E003}) rejects any commit that
touches anything else, so one task can never corrupt another's data.

\paragraph{2. Immutability with a corrections overlay.}
Motivated by the same incident: the repair was a separate downstream
task that restored 48 CSV files across 34 feature sets without
rewriting the original step's record. Completed task folders are
immutable --- a verifier rejects any branch that modifies a merged
task's files; downstream tasks fix mistakes via correction files in
their own folders, applied at read time by aggregators.

\paragraph{3. Aggregators-only cross-task reading.}
Motivated by F07, in which a hand-maintained tracker file reported
\texttt{Completed Tasks (75 of 64)} --- an impossible count produced
by manual editing during a busy week. Skills must read through
\path{arf/scripts/aggregators/}, never by walking task folders
directly.

\paragraph{4. Materialised project overview for human observability.}
Derived from the same observability problem (F07) and from two
further incidents (F08, F09) in which stale or missing per-feature
metrics misled human reviewers. A dedicated
\path{arf/scripts/overview/materialize.py} regenerates
\path{overview/} --- a committed, browsable, GitHub-renderable
dashboard of all aggregator output. The materialisation is
automatic, the artefact is static and reviewable, and
\path{overview/} is the only human-facing summary in the repository.

\paragraph{5. Spec-verified artefacts.}
Motivated by F01 and F04 (target leakage in generated feature code).
Every produced artefact has a versioned specification and a
corresponding verifier that checks it before commit. We elaborate
this principle in \S\ref{sec:system-verificators}.

\paragraph{6. Comprehensive logging.}
Motivated by debugging pain across the campaign and bounded by F03 as
a limits example. Every CLI invocation in a task branch is wrapped in
\path{arf/scripts/utils/run_with_logs.py}, which records the command,
stdout, stderr, exit code, and timestamp; a verifier refuses to
mark a step complete if an expected command log is missing, so the
work that happened and the record of it cannot diverge.

\paragraph{7. Subagent isolation.}
Motivated by context-window degradation in long-horizon agents.
Complex tasks run as a chain of subagents (research, planning,
implementation, analysis, reporting); each subagent has its own
context and sees only the inputs it needs.

\begin{figure}[t]
\centering
\begin{tikzpicture}[
  stage/.style={
    draw, rounded corners=2pt, thick,
    minimum width=2.12cm, minimum height=0.76cm,
    align=center, font=\scriptsize
  },
  flow/.style={-Latex, thick, draw=gray!65},
]
  \node[stage] (s1) at (0,2.4)
    {\textbf{1. Papers}\\[-1pt]{\tiny\itshape\color{black!60}read/summarize}};
  \node[stage] (s2) at (2.55,2.4)
    {\textbf{2. Internet}\\[-1pt]{\tiny\itshape\color{black!60}search/download}};
  \node[stage] (s3) at (5.10,2.4)
    {\textbf{3. Code}\\[-1pt]{\tiny\itshape\color{black!60}prior tasks/libs}};
  \node[stage] (s6) at (0,1.18)
    {\textbf{6. Analysis}\\[-1pt]{\tiny\itshape\color{black!60}metrics/charts}};
  \node[stage] (s5) at (2.55,1.18)
    {\textbf{5. Implement}\\[-1pt]{\tiny\itshape\color{black!60}code/train/eval}};
  \node[stage] (s4) at (5.10,1.18)
    {\textbf{4. Planning}\\[-1pt]{\tiny\itshape\color{black!60}approach/budget}};
  \node[stage] (s7) at (0,-0.04)
    {\textbf{7. Lit. compare}\\[-1pt]{\tiny\itshape\color{black!60}published work}};
  \node[stage] (s8) at (2.55,-0.04)
    {\textbf{8. Reporting}\\[-1pt]{\tiny\itshape\color{black!60}results/suggestions}};
  \node[stage] (s9) at (5.10,-0.04)
    {\textbf{9. PR \& merge}\\[-1pt]{\tiny\itshape\color{black!60}review/correct}};

  \draw[flow] (s1.east) -- (s2.west);
  \draw[flow] (s2.east) -- (s3.west);
  \draw[flow] (s3.south) -- (s4.north);
  \draw[flow] (s4.west) -- (s5.east);
  \draw[flow] (s5.west) -- (s6.east);
  \draw[flow] (s6.south) -- (s7.north);
  \draw[flow] (s7.east) -- (s8.west);
  \draw[flow] (s8.east) -- (s9.west);

  \node[font=\scriptsize, text=black!75] at (2.55,-0.72)
    {logged \(\cdot\) spec-verified \(\cdot\) corrected, not rewritten};
\end{tikzpicture}
\caption{The nine-step lifecycle every task passes through: papers
$\rightarrow$ internet $\rightarrow$ code $\rightarrow$ planning
$\rightarrow$ implementation $\rightarrow$ analysis $\rightarrow$
literature comparison $\rightarrow$ reporting $\rightarrow$ PR \& merge.
In complex tasks each step runs in its own subagent and writes to a
known place inside the task folder.}
\label{fig:lifecycle}
\end{figure}
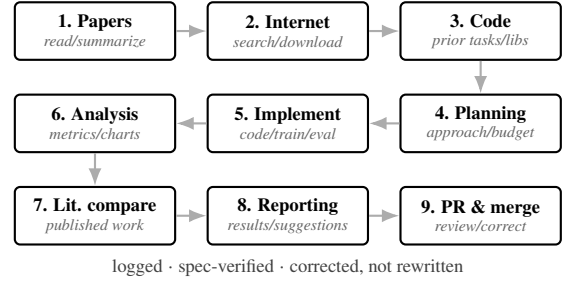

Figure~\ref{fig:lifecycle} shows the lifecycle every task passes
through; \S\S\ref{sec:system-verificators}--\ref{sec:system-subagents}
describe how the principles above are enforced in practice.

\subsection{Verifiers: enforcing the principles}
\label{sec:system-verificators}

Rules are only rules if they are followed; structure is only real if
something checks it. Glite ARF's verifier layer turns the
principles into deterministic Python scripts that fail loudly when
violated, so an agent cannot silently drift past a rule by claiming to
have followed it. Our companion BEA~2026 system paper describes the
same mechanism in its framework-overview section \citep{glite2026bea}:
``task isolation (one directory, one branch, one pull request per
experiment), artefact specifications checked by verifiers before
commit, and an immutable log of every experiment step.''

Every artefact has a versioned specification. Framework-level
artefacts (task folders, results files, logs) live under
\path{arf/specifications/}. Asset-type artefacts (papers, datasets,
models, predictions, answers, suggestions, libraries) live under
\path{meta/asset_types/<kind>/specification.md}. Specifications are
plain markdown with a numeric version field; files produced under a
spec carry a matching \texttt{spec\_version} field, so format
evolution is auditable rather than silent, and verifiers validate
each artefact against the version it was written under.

Each specification has a verifier script under
\path{arf/scripts/verificators/} that parses the artefact and produces
error or warning diagnostics with stable codes (e.g.\
\texttt{FD-E001}: a mandatory task subdirectory is missing;
\texttt{PM-E003}: the task branch modified a file outside its folder).
Errors block commit and merge --- enforced both by the
prestep/poststep lifecycle gate and as required pull-request checks ---
while warnings surface concerns without blocking. Verifiers are
deliberately allowed to be noisy and duplicated: two verifiers
checking the same thing from different angles catch cases either alone
might miss.

What the verifiers deliberately do \emph{not} judge is semantic
validity --- whether an experiment is well-designed or a baseline
appropriate. That judgement stays with the human researcher (role 1,
\S\ref{sec:lessons}); the framework's job is to make the structural
substrate trustworthy so that human judgement operates on a reliable
record.

\subsection{Aggregators and the corrections overlay}
\label{sec:system-aggregators}

Each datum has a single home --- the task folder that produced it.
When the framework needs a combined view (every paper across every
task, every metric filtered by category, every cost summed across the
campaign), an aggregator script in \path{arf/scripts/aggregators/}
walks \path{tasks/}, applies filters and the corrections overlay, and
returns the canonical answer. Snapshot output is committed to
\path{overview/} (principle 4) so humans can browse the materialised
view on GitHub without rerunning the scripts.

Completed task folders are immutable. When a downstream task discovers
an earlier result was wrong --- a paper misclassified, a metric
mis-aggregated, a feature later determined to leak the target --- it
does not reach back into the earlier task's folder. It writes a small
correction file in its own \path{corrections/} directory, and
aggregators apply the correction overlay at read time, returning the
effective view that incorporates every later fix while preserving the
original record.

This discipline addresses one of the campaign's most concrete failure
modes (F07): a hand-maintained \path{tracker/completed.md} reporting
\texttt{Completed Tasks (75 of 64)} --- an impossible count produced
by manual editing during a busy week. With aggregators replacing
manual cross-task summaries, the canonical answer is recomputable and
traceable.

\subsection{Subagent isolation and parallelism}
\label{sec:system-subagents}

Complex tasks run as a chain of subagents --- research, planning,
implementation, analysis, reporting --- each with its own context
window and inputs scoped to what it needs. A research subagent that
has read fifty paper summaries does not pollute the planning
subagent that follows. This addresses a hard limit of current LLMs:
they degrade as context fills up. Splitting the work into
context-bounded stages keeps each one focused and gives the framework
a natural insertion point for verification between stages.

\begin{figure}[t]
\centering
\includegraphics[width=\columnwidth]{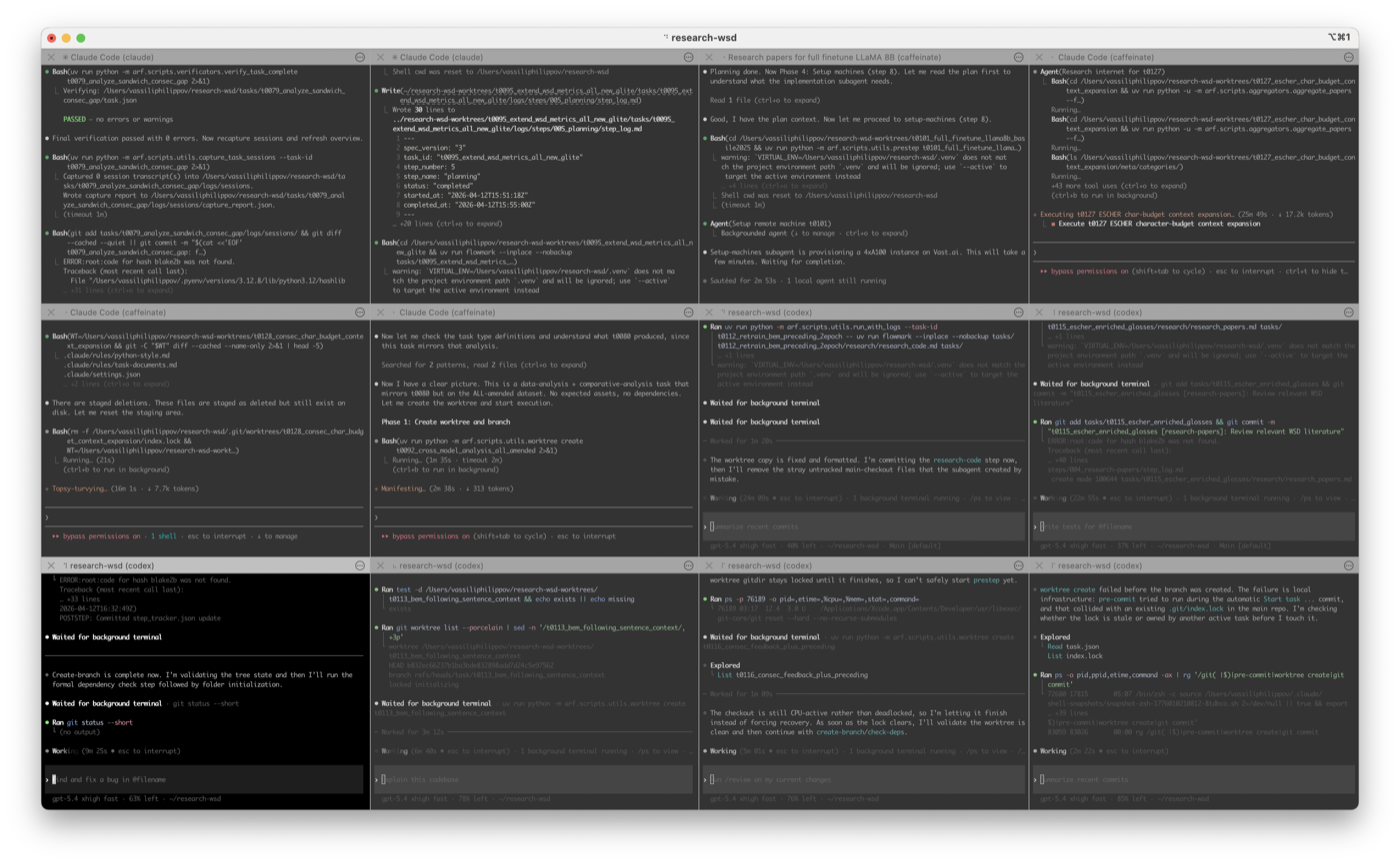}
\caption{Twelve agent sessions running in parallel on a single 48~GB
Mac during the BEA~2026 campaign; each tile is one
\texttt{task/<task\_id>} worktree. Figure~\ref{fig:concurrency}
shows the analogous concurrency profile, measured over time, for the
WSD campaign (\S\ref{sec:in-use}).}
\label{fig:twelve-agents}
\end{figure}

Tasks run in parallel by living in separate git worktrees on separate
branches (\texttt{task/<task\_id>}). The researcher opens a
coding-agent session per task; ARF supplies the worktree and branch
conventions, the verifier gates, and the merge discipline that let
many sessions run at once without interfering. Up to twelve sessions
ran simultaneously on a single 48~GB Mac during the BEA~2026 campaign
(Figure~\ref{fig:twelve-agents}), and no merge conflict reached
\texttt{main}. The pattern matches concurrent prior work --- EvoSkill
\citep{alzubi2026evoskill} frames each agent program as a git branch;
CORAL \citep{qu2026coral} describes ``isolated workspaces while
sharing access to the same evaluator and shared persistent memory'' ---
but in ARF the parallelism is at the \emph{task} granularity inside a
\emph{campaign}, not at the agent granularity inside a single
optimisation problem. Section~\ref{sec:in-use} quantifies this
concurrency across campaigns.

\section{Case Study: BEA 2026}
\label{sec:case-study}

This paper's contribution is the framework; the companion BEA~2026
system paper \citep{glite2026bea} describes the prediction system, and
we summarise its results here only as the setting in which the
framework was exercised.

\subsection{The shared task}

BEA~2026 \citep{skidmore2025kvl} introduced an L1-aware
vocabulary-difficulty prediction task: given an English target word, a
partial-spelling clue, an L1 translation, and a context sentence in
the learner's L1 (Spanish, German, or Mandarin), predict the word's
psychometric difficulty as a continuous, GLMM-calibrated score.
Labels are derived from roughly 3.3M test responses produced by
$100{,}000{+}$ test-takers on the British Council Knowledge-based
Vocabulary Lists \citep{schmitt2024kvlcorpus}. The task scores by
RMSE (lower is better) and reports Pearson correlation as a secondary
metric. The closed track forbids generative LLMs, paid APIs, and
additional training data; the open track has no such restriction.

\subsection{The campaign}

We developed our submission inside Glite ARF over a multi-week
research campaign. The campaign comprised 273 tracked task folders
(146 of type \texttt{experiment-run}; 247 completed, 9 permanently
failed, 17 cancelled --- App~\ref{app:aggregators}) across 129 feature
sets totalling 1{,}161 numeric feature columns organised into seven
domain families plus a derived bucket. Every feature CSV, every
fold-level score, every commit message is visible in the project's PR
history. Up to twelve agent sessions ran in parallel, orchestrated
from a single 48~GB Mac, with rented A100 instances used only for
scaled-encoder fine-tuning and decoder-LLM LoRA training. Third-party
cost was \$449.69 in LLM API spend for feature engineering and model training (Anthropic \$287.12, OpenAI
\$162.57) plus \$48.62 in rented A100 compute --- \$498.31 total --- and
${\sim}100$ wall-hours of local compute (App~\ref{app:costs}).
Aggregator snapshot excerpts appear in App~\ref{app:aggregators}.

\subsection{Headline result}

\begin{figure}[t]
\centering
\includegraphics[width=\columnwidth]{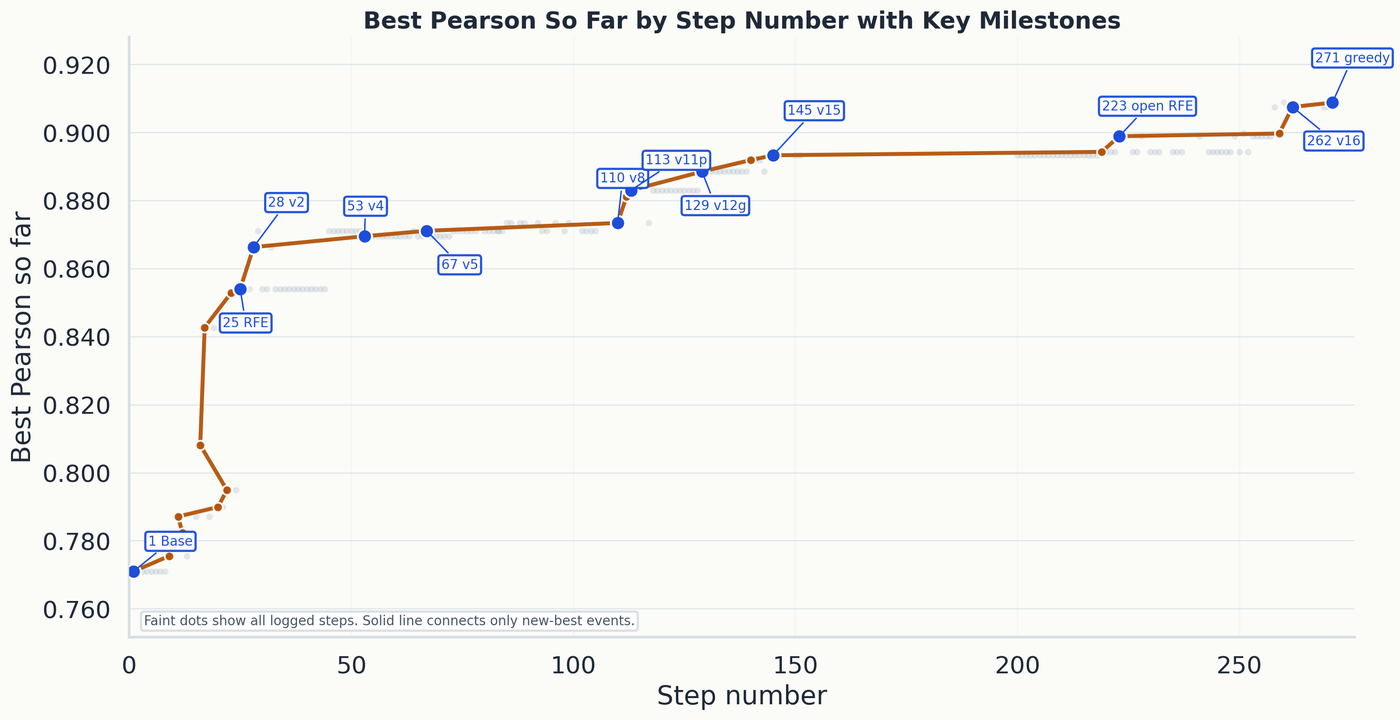}
\caption{Best dev-set Pearson correlation across the BEA~2026
campaign, by step index (one step per tracked task). Most tracked
tasks leave the best-so-far value unchanged; a few produce a step.}
\label{fig:pearson-staircase}
\end{figure}

Our system placed first in the closed track on all three L1s and
second in the open track on all three L1s
(Table~\ref{tab:leaderboard}). The average RMSE reduction over the
official baseline was 29.9\% closed-track (28.2\%~ES, 29.6\%~DE,
31.9\%~CN) and 35.9\% open-track (37.1\%~ES, 34.5\%~DE, 36.2\%~CN).
The best single component is a LLaMA-3.1-8B LoRA regression head with
a K-fold RMSE of 0.831 at 0.13\% trainable parameters.
Figure~\ref{fig:pearson-staircase} shows best-so-far dev-set Pearson
across the campaign --- from 0.78 at task \texttt{t0001} to 0.91 by
task \texttt{t0270}, with most experiments producing no improvement
and a small number producing the large jumps that give the staircase
its shape. We show this only to illustrate the search dynamics: a
best-so-far curve is monotonic by construction, Pearson is the
secondary metric (the official objective is RMSE), and task index is
not wall-clock time in a parallel campaign.

\begin{table*}[t]
\small
\centering
\begin{tabular}{llrrrr}
\toprule
Track & System & ES (RMSE) & DE (RMSE) & CN (RMSE) & Avg \\
\midrule
Closed & \textbf{1. Ours}              & \textbf{0.903} & \textbf{0.885} & \textbf{0.776} & \textbf{0.855} \\
Closed & 2. Best non-Glite, per L1\textsuperscript{\textdagger}    & 0.975 & 0.903 & 0.816 & --- \\
Closed & Baseline                      & 1.257 & 1.258 & 1.140 & 1.218 \\
\midrule
Open   & 1. Sakura                     & 0.742 & 0.723 & 0.630 & 0.698 \\
Open   & \textbf{2. Ours}              & \textbf{0.754} & \textbf{0.764} & \textbf{0.660} & \textbf{0.726} \\
Open   & 3. TeamXBC                    & 0.876 & 0.826 & 0.722 & 0.808 \\
Open   & Baseline                      & 1.198 & 1.166 & 1.034 & 1.133 \\
\bottomrule
\end{tabular}
\caption{Official BEA~2026 test-set leaderboard (RMSE, lower is
better). \textsuperscript{\textdagger}\,The closed-track row is the
best \emph{non-Glite} result \emph{per L1} --- \textit{uogal} for ES
and DE, \textit{Sakura} for CN --- and is not a single system, so we
omit a cross-L1 average for it. Open-track rows are single teams.}
\label{tab:leaderboard}
\end{table*}

\subsection{Structured provenance caught the leakage}
\label{sec:case-study-leakage}

The campaign's sharpest demonstration of verifier-driven research
is the leakage post-mortem reported in our BEA system paper
\citep{glite2026bea}. An intermediate ensemble (v13) scored an
implausibly strong K-fold RMSE of 0.609; because every feature CSV
carried a versioned specification and every fold-level score was
traceable to the code revision that produced it, we localised the
cause within minutes: four feature sets were leaking the target
variable.
\texttt{outlier\_surgery} passed the GLMM target into LLM prompts as
``actual difficulty.'' \texttt{cross\_l1\_oof} used cross-L1
out-of-fold predictions with an item-id split shared across L1s.
\texttt{heteroscedastic} regressed on GLMM-score residuals.
\texttt{annotation\_noise} computed noise statistics from the GLMM
target. Each used the GLMM target that is available at
training/cross-validation time but not for hidden-test items, which is
precisely why they leaked. All four were quarantined into a
\path{removed_leaking_features/} folder via the corrections overlay,
and the rebuilt v14 ensemble scored RMSE~0.802 --- the corrected
estimate. The structural layer cannot judge that a feature is
\emph{semantically} leaking --- that call was a human's --- but the
per-feature specifications and per-fold provenance turned what could
have been a silent contaminated submission into a fix made in minutes
and fully auditable after the fact. This is verifier-driven
research working as intended: structure makes the failure findable and
the correction trustworthy.

\subsection{Compliance audit at scale}

The closed track's restrictions (no generative LLMs, no paid APIs, no
extra training data, no cross-L1 training) created a second governance
problem. These restrictions apply to the \emph{submitted system} ---
its features and inference path --- not to development assistance, so
using coding agents to author and run pipeline code is permitted while
LLM-derived features are not; keeping that line required per-feature
accounting. We developed a per-column compliance schema that records,
for each feature column, the external models, APIs, datasets, and
computation paths used to produce it, detailed in that paper's
compliance-engineering section \citep{glite2026bea}.
When the BEA organisers issued five rule clarifications in March
2026, the per-column metadata enabled a finer-grained audit than
feature-set-level checks would have: 17 features were reclassified
after the clarifications and our internal audit added two more,
leaving 57 of 129 feature sets (249 of 1{,}161 columns)
closed-track-eligible --- a 56\% rejection rate that forced a full
pipeline rebuild in the final week. We submitted a 469-line LLM-usage
disclosure with the closed-track system. The rebuilt closed-track
system retained its 1st-place ranking on every L1.

\section{The framework in use}
\label{sec:in-use}

To characterise Glite ARF independently of any single result, we mined
the logs of two further ARF campaigns in different
domains: \texttt{research-wsd} (word-sense disambiguation; 167
completed tasks over 74 days) and \texttt{research-ace-cefr} (English
CEFR readability). Figures below derive from each campaign's logs
(\texttt{research-ace-cefr} is public; the \texttt{research-wsd} logs
are author-held). Together with BEA, this is
three multi-week campaigns in
three domains driven by the same framework code
(Table~\ref{tab:campaign-metrics}).

\begin{table}[t]
\centering\small
\begin{tabular}{@{}lrrr@{}}
\toprule
 & WSD & Ace-CEFR & BEA \\
\midrule
Completed tasks      & 167 & 35 & 247 \\
Experiment runs      & 72  & 15 & 146 \\
Merged branches      & 271 & 34 & --- \\
Logged commands      & 17{,}915 & 1{,}916 & --- \\
Peak concurrency     & 12  & 3  & 12 \\
Spend (USD)          & 4{,}039 & 19 & 498 \\
\bottomrule
\end{tabular}
\caption{Three ARF campaigns in three domains on identical framework
code. WSD and Ace-CEFR are mined from their campaign logs (Ace-CEFR
public; WSD author-held); BEA is from \S\ref{sec:case-study}.}
\label{tab:campaign-metrics}
\end{table}

\begin{figure}[t]
\centering
\includegraphics[width=0.92\columnwidth]{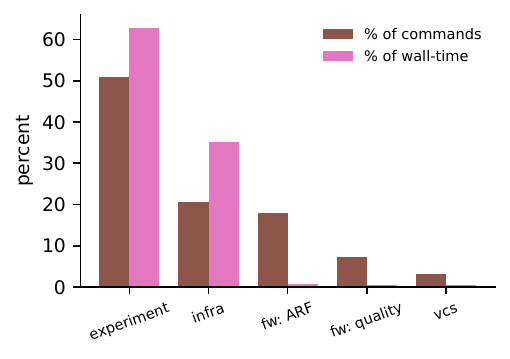}
\caption{Command-level overhead (WSD): ARF's own scripts are
${\sim}25\%$ of commands but ${\sim}1\%$ of wall-clock time.}
\label{fig:overhead-bars}
\end{figure}

\paragraph{The structural machinery is cheap.}
Classifying the 17{,}915 logged WSD commands, the framework's own
scripts --- verifiers, aggregators, \path{run_with_logs}, and the
lint/type/test gate --- are about a quarter of all commands but only
${\sim}1\%$ of wall-clock time (Figure~\ref{fig:overhead-bars});
experiment work dominates runtime. Per-stage timing agrees: the
framework-administrative steps (branch creation, dependency checks,
folder init) take seconds, while implementation steps take tens of
minutes. Verifier-driven research buys its guarantees at negligible
time cost.

\begin{figure}[t]
\centering
\includegraphics[width=0.92\columnwidth]{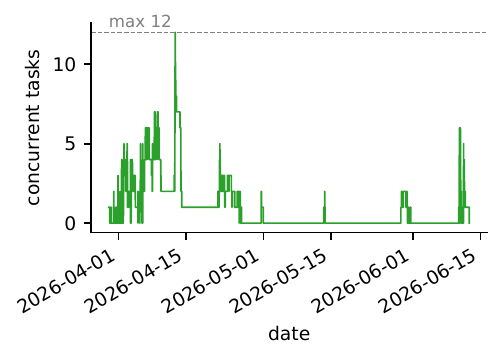}
\caption{Concurrent in-flight WSD tasks over the 74-day campaign;
peak twelve.}
\label{fig:concurrency}
\end{figure}

\paragraph{Parallelism, measured.}
Reconstructing in-flight tasks from their timestamps, the WSD campaign
sustained bursts of concurrent work peaking at twelve simultaneous
tasks (Figure~\ref{fig:concurrency}), confirming the ``up to twelve
agents'' figure of \S\ref{sec:system-subagents} with data rather than
a screenshot. The profile is bursty, matching a human driving role 1
while many task agents run underneath.

\paragraph{Isolation and portability hold across domains.}
Across 305 merges in the two mined campaigns, no task corrupted
another task's data or the shared \texttt{main} branch, and no merge
conflict reached \texttt{main}. The three campaigns span word-sense
disambiguation, readability, and vocabulary difficulty, yet the
framework code under \path{arf/} is byte-for-byte identical: only
\path{project/}, \path{tasks/}, and \path{meta/} differ. The framework
transfers across domains without modification.

\section{Lessons and Limits}
\label{sec:lessons}

\paragraph{The structural layer is cheap.}
A natural worry is that wrapping every command, verifying every
artefact, and re-reading specifications at each step is expensive. It
is not: across the campaigns the framework's own machinery accounts for
about a quarter of all commands but only ${\sim}1\%$ of wall-clock time
(\S\ref{sec:in-use}). The BEA campaign cost approximately \$450 in LLM
API spend (\$498 total third-party) across 273 tracked tasks --- low
per experiment. The overhead is intentional and well spent: when an
agent in week eight needs to know what an agent in week one actually
did, the answer comes from a recorded, verified artefact, not a guess.

\paragraph{Human-in-the-loop is intentional, not interim.}
Role 1 in our three-role stack --- choosing which hypotheses to test
next --- is human-driven by design, not because we are waiting for
stronger models. The METR time-horizon study
\citep{metr2025horizon} measures the task length at which a model
succeeds half the time at roughly one hour for the strongest model it
evaluated, with the more reliable 80\% horizon far shorter; a research
campaign runs for weeks. The AI Scientist v1 \citep{lu2024sakana} was
evaluated by \citet{beel2025evaluating}, which found that 42\% of its
generated experiments failed due to coding errors --- exactly the
failure mode that emerges when the strategic-decision role is
delegated to current models. We make a different bet: keep role 1
human, delegate role 2 entirely, and put deterministic Python scripts
on role 3.

\paragraph{What verifiers do not catch.}
Verifiers are deterministic; they catch structural errors
--- files outside the task folder modified, logs missing, JSON
malformed against its spec, fold-level scores untraceable to a code
revision. They do not catch semantic errors --- a wrong analysis, a
wrong baseline, a methodologically inappropriate but plausible-looking
experimental design. \citet{schmidgall2025agentlab} report that
automated paper-quality evaluation overestimates human reviewer
scores by ${\sim}2$ points; we expect the same gap in any
verifier-style evaluation of scientific merit. Glite ARF's
contribution is that it makes a class of structural failures
\emph{detectable and auditable} while leaving semantic judgement where
it belongs --- with the human researcher (role 1).

\paragraph{Limitations.}
Two further limits are worth stating plainly. First, ARF assumes a
\emph{single} human operator: its task-index and merge conventions are
not designed for several people driving role~1 concurrently, which
would risk index collisions and duplicated work. Second, the
verifiers are not agent-proof --- they are ordinary scripts in the
repository, so an agent explicitly instructed to bypass a rule can
weaken or rewrite the verifier that checks it. ARF targets
accidental agent drift in a non-adversarial setting, not a determined
or adversarial agent.

\section{Availability}
\label{sec:availability}

Glite ARF \citep{glite2026arf} is released under the Apache~2.0
licence (version~0.1.0) at
\url{https://github.com/GliteTech/glite-arf}. A public ARF
demo project accompanies it: English CEFR readability
(\url{https://github.com/GliteTech/research-ace-cefr}). The
framework code under \path{arf/} is identical across all three
campaigns; only \path{project/}, \path{tasks/}, and \path{meta/} carry
project-specific content (\S\ref{sec:in-use}).
Appendix~\ref{app:failure-modes} summarises the failure modes that
shaped the design.

\paragraph{Future work.}
We plan automatic citation verification (motivated by F05), per-column
compliance audits as a reusable methodological contribution,
heartbeat-style periodic agent reflection adapted from CORAL
\citep{qu2026coral}, and a controlled study --- seeding the
Appendix~\ref{app:failure-modes} incidents into test repositories ---
to measure verifier detection rates directly. Replacing role 1 with
an autonomous brainstorming agent is not a near-term priority --- see
above.

\section*{Ethics Statement}
\label{sec:ethics}

Glite ARF runs LLM coding agents (Claude Code, Codex CLI)
autonomously on the operator's machine, with shell execution, file
writes, \texttt{git push} permissions, and paid API calls. The
framework's verifier layer enforces structural integrity, but it is
not a security sandbox: it does not constrain what an agent's shell
commands can do to the host machine. The repository ships a permissive
\path{.claude/settings.json} suitable for research environments and
documents the security trade-offs in
\path{arf/docs/explanation/safety.md};
operators on machines with sensitive data should narrow the
allow-list before running.

The BEA~2026 dataset is distributed under the British Council's
Knowledge-based Vocabulary Lists licence
\citep{schmitt2024kvlcorpus}; we used it under the shared-task
agreement. No personally identifying information appears in any
artefact released with this paper. LLM API spend for the BEA campaign
was approximately \$450 (\$498 total third-party cost including rented
compute) and is itemised in Appendix~\ref{app:costs}. Our campaign ran
predominantly on a 48~GB Mac (CPU/GPU) and on rented A100 GPU hours;
we did not measure carbon emissions and therefore do not characterise
the campaign's footprint.

The framework lowers the cost of running multi-agent research
campaigns. We acknowledge that the same lowered cost could be used
to run campaigns whose hypotheses are not socially beneficial; the
framework imposes structure on \emph{how} research is conducted, not
on \emph{what} is researched. Role 1 in the three-role stack
(hypothesis selection) remains human, in part for the structural
reasons argued in \S\ref{sec:lessons} and in part because the human
is where ethical accountability lives.

\bibliography{refs}

\appendix

\section{Task folder structure}
\label{app:task-folder}

Every ARF task lives in \path{tasks/tNNNN_<slug>/} on a branch called
\texttt{task/<task\_id>}. The folder follows a fixed layout
(Listing~\ref{lst:task-folder}); the verifier
\path{verify_task_folder.py} (prefix \texttt{FD}) rejects any commit
that adds a top-level file outside the listed set or omits a mandatory
subdirectory.

\begin{lstlisting}[caption={On-disk layout of a single ARF task
folder.},label={lst:task-folder},frame=single]
tasks/t0042_example_task/
  task.json, step_tracker.json
  plan/, research/
  assets/{papers,datasets,libraries,answers,
          suggestions,models,predictions}/
  results/{summaries,metrics,costs,images}/
  corrections/, intervention/
  logs/{commands,steps,searches,sessions}/
\end{lstlisting}

The public PR for the corresponding worked example
(\path{tasks/t0042_example_task/}) is linked from the framework
README. Reviewers can browse the actual task folder there to see
what each subdirectory contains in practice.

\section{Verifier catalogue and log spec}
\label{app:verificators}

\subsection{Verifier scripts}

Verifiers live under \path{arf/scripts/verificators/}. Each script
parses a class of artefact and emits diagnostic codes of the form
\texttt{<2-CHAR>-<E|W>NNN}. Errors block commit and merge (via the
prestep/poststep gate and required pull-request checks); warnings
surface concerns without blocking. A representative subset of the
shipped scripts:

\begin{itemize}[leftmargin=*,itemsep=0pt,topsep=2pt,parsep=0pt]
\item \texttt{verify\_task\_folder.py} (\texttt{FD}) --- structural
  integrity of \path{tasks/<id>/}: the mandatory subdirectory layout
  and required files.
\item \texttt{verify\_task\_results.py} (\texttt{TR}) and
  \texttt{verify\_metrics.py} (\texttt{MT}) --- presence and format of
  \path{results/} files (\texttt{metrics.json},
  \texttt{results\_summary.md}); cross-checks the step tracker against
  produced artefacts.
\item \texttt{verify\_logs.py} (\texttt{LG}) --- log presence and
  structure against \path{arf/specifications/logs_specification.md}.
\item \texttt{verify\_pr\_premerge.py} (\texttt{PM}) --- the pre-merge
  gate; \texttt{PM-E003} rejects a commit that modifies files outside
  the task folder beyond the allow-list, and the immutability check
  rejects edits to merged tasks.
\item \texttt{verify\_corrections.py} (\texttt{CR}) --- every
  correction file targets an existing earlier task and follows the
  correction schema.
\item Asset-type verifiers under
  \path{meta/asset_types/<kind>/verificator.py} --- per-asset-type
  specifications for papers, datasets, models, predictions, answers,
  suggestions, and libraries.
\end{itemize}

Every specification has a corresponding verifier, so each artefact
class an agent can produce is checked before it is committed.

\subsection{Per-step log format}

Logs live under the task's \path{logs/} directory in four mandatory
subdirectories (logs specification v5), summarised in
Table~\ref{tab:log-format}.

\begin{table}[t]
\small
\centering
\begin{tabular}{ll}
\toprule
Subdirectory & Contents \\
\midrule
\texttt{commands/} & one JSON record per CLI invocation \\
                   & (\texttt{command}, \texttt{exit\_code}, \\
                   & \texttt{timestamp}) plus \texttt{.stdout.txt} \\
                   & and \texttt{.stderr.txt} side-files \\
\texttt{steps/}    & one folder per step; \texttt{step\_log.md} \\
                   & with YAML frontmatter \\
\texttt{searches/} & one record per internet search query \\
\texttt{sessions/} & agent session-capture outputs \\
\bottomrule
\end{tabular}
\caption{Log directory layout (specification v5). Command logs are
auto-generated by \texttt{run\_with\_logs.py}, which wraps every CLI
invocation in a task branch.}
\label{tab:log-format}
\end{table}

\section{Cost breakdown}
\label{app:costs}

Total third-party cost for the BEA~2026 campaign was \$498.31, of
which LLM API spend was \$449.69 (Anthropic \$287.12, OpenAI \$162.57)
and rented A100 compute (Vast.ai) was \$48.62; on top of this the
campaign used ${\sim}100$ wall-hours of local compute on a 48~GB Mac
(see App~\ref{app:aggregators} for the aggregator snapshot these
figures come from). Table~\ref{tab:llm-cost-top} reports the
highest-LLM-cost feature experiments, drawn from the campaign's
per-task cost aggregator (App~\ref{app:aggregators}).

\begin{table}[t]
\small
\centering
\setlength{\tabcolsep}{3.8pt}
\begin{tabular}{@{}lrr@{}}
\toprule
Feature experiment & Cost (\$) & $\Delta$RMSE \\
\midrule
LLM rubric (4o-mini, full pool)  & 1.93 & $-0.015$ \\
LLM rubric (4o-mini, ablation)   & 9.72 & $-0.012$ \\
LLM-WSD (\path{gpt-4o})          & 36.14 & $-0.008$ \\
Polysemy contrast (LLM)          & 22.47 & $-0.004$ \\
Counterfactual cue sensitivity   & 41.38 & $-0.011$ \\
Annotation-noise LLM (leaking) & 14.52 & ---\textsuperscript{\textdagger} \\
\bottomrule
\end{tabular}
\caption{Highest-cost LLM-API feature experiments and their effect
on full-system K-fold RMSE. \textsuperscript{\textdagger}\,The
\texttt{annotation\_noise} feature was one of the four leaking
sets quarantined by the audit described in
\S\ref{sec:case-study-leakage} and was excluded from the final
ensemble.}
\label{tab:llm-cost-top}
\end{table}

We distinguish two cost categories.

\paragraph{Per-experiment cost.}
The cost of running a feature experiment itself --- LLM-rubric
queries, decoder-LLM LoRA training calls, paid embedding lookups.
This is what Table~\ref{tab:llm-cost-top} reports and what drives the
\$449.69 LLM API total. These figures cover paid LLM \emph{API} calls
for feature engineering and model training only; the Claude Code and
Codex coding agents that authored and ran the pipeline operated under
flat-rate subscriptions during the campaign, so their token usage
carried no per-task marginal cost and is not part of the \$449.69.
Teams paying metered per-token rates for the agents should budget for
that separately.

\paragraph{Framework overhead.}
The structural machinery --- \texttt{run\_with\_logs} command wrapping,
spec checks, verifier runs, aggregator materialisation, and the
subagent pipeline --- is cheap in wall-clock terms: across campaigns it
is about a quarter of all commands but only ${\sim}1\%$ of wall-clock
time (\S\ref{sec:in-use}, Fig.~\ref{fig:overhead-bars}). The added
token cost per task is modest relative to the experiment work itself,
and is the intended price of a fully recorded, verifiable research
trail.

\section{Observed failure modes}
\label{app:failure-modes}

Table~\ref{tab:failure-modes} summarises five representative failure
classes observed during the BEA~2026 campaign and the framework
principle each one motivated
(App~\ref{app:aggregators} excerpts the campaign snapshots). Several
incidents occurred under earlier versions of the framework and
motivated specific later additions; the per-feature-CSV
specifications added in response to F01 are what later made the
leakage in \S\ref{sec:case-study-leakage} \emph{auditable} --- the
audit itself was human-driven, not triggered by a verifier.

\par\smallskip
\refstepcounter{table}\label{tab:failure-modes}
\begin{center}
{\footnotesize
\setlength{\tabcolsep}{2.5pt}\renewcommand{\arraystretch}{1.12}
\begin{tabular}{@{}p{1.7cm}p{3.45cm}p{2.15cm}@{}}
\toprule
Class (ref.) & Concrete example & Motivated principle \\
\midrule
Data corruption (F02) & Step 146 reran train+dev+test; 38 feature sets
corrupted & Task isolation; immutability \\
Semantic invalidity (F01, F04) & \texttt{outlier\_surgery} leaked
target labels; v13 RMSE 0.609 $\rightarrow$ v14 0.802 &
Spec-verified artefacts; corrections overlay \\
Stale summaries (F07) & \texttt{Completed Tasks (75 of 64)} from manual
tracker editing & Aggregators-only reading; materialised overview \\
Citation hallucination (F05) & RemBERT BibTeX had 3 of 5 coauthors
hallucinated & Future citation verifier \\
Resource state drift (F06) & A100 status left as ``Running step 323''
after no result report & Future teardown verifier \\
\bottomrule
\end{tabular}
\par}
\end{center}
\vspace{0pt}
\noindent\small Table~\thetable: Five representative failure modes and the framework
principles each one motivated. References in parentheses resolve to
the full incident catalogue.
\par\medskip\normalsize

We adopt the deliberately bounded language ``motivated'' rather than
``prevented'' or ``solved.'' Verifiers catch a class of
structural error; semantic errors such as F01 and F04 (target
leakage in generated feature code) become \emph{auditable} once the
artefact is spec-verified, but the audit itself is run by a
human. \S\ref{sec:lessons} returns to this distinction.

\section{Secondary case study: Ace-CEFR}
\label{app:ace-cefr}

\par\smallskip
\refstepcounter{table}\label{tab:ace-cefr-task-types}
\begin{center}
\begin{tabular}{@{}lr@{}}
\toprule
Task type & Tasks \\
\midrule
\texttt{experiment-run}        & 11 \\
\texttt{build-model}           &  6 \\
\texttt{baseline-evaluation}   &  4 \\
\texttt{download-paper}        &  3 \\
\texttt{comparative-analysis}  &  3 \\
\texttt{feature-engineering}   &  2 \\
\texttt{download-dataset}      &  2 \\
\texttt{literature-survey}     &  2 \\
\texttt{data-analysis}         &  2 \\
others (8 task types)          &  3 \\
\midrule
\textbf{Total}                  & \textbf{38} \\
\bottomrule
\end{tabular}
\end{center}
\vspace{0pt}
\noindent\small Table~\thetable: Public Ace-CEFR task-type breakdown.
All counts use generic ARF template types; the project adds no
project-specific task types. These counts assign each task a single
type and sum to~38; the \texttt{experiment-run} figure in
Table~\ref{tab:campaign-metrics} (15) instead counts every task
\emph{tagged} \texttt{experiment-run}, as a task may carry several
types.
\par\medskip\normalsize

We illustrate reuse with a public ARF project in a related
domain. \texttt{research-ace-cefr} \citep{glite2026acecefr} is
author-run and still in progress, so it shows reuse rather than
independent adoption. It targets English CEFR readability on the
Ace-CEFR benchmark \citep{kogan2025acecefr}: 890 conversational
passages labelled on the CEFR 1--6 scale, split 445 train / 445 test.
The project aims to beat the published BERT~+~PaLM ensemble baseline
(0.33 MSE) under a stratified split that does not tune on test.

The snapshot contains 38 task folders across 17 task types
(Table~\ref{tab:ace-cefr-task-types}); 35 have executed and 3 are
queued as planned next steps. The materialised \path{overview/}
dashboard is committed to the repository, so readers can browse
metrics, per-task status, and corrections on GitHub without rerunning
code. Ace-CEFR shares no project-specific code with the BEA campaign:
\path{arf/} is identical, while \path{project/}, \path{tasks/}, and
\path{meta/} carry project-specific content.

\section{Aggregator output snapshots}
\label{app:aggregators}

Aggregator output snapshots were taken on 2026-05-06;
Listing~\ref{lst:agg-snapshots} shows representative excerpts.
Every quantitative claim in
\S\ref{sec:case-study} traces to one of these files.

\begin{lstlisting}[caption={Aggregator snapshot excerpts.},
label={lst:agg-snapshots},
frame=single,
basicstyle=\ttfamily\scriptsize]
date=2026-05-06; tasks=273/247/9/17
types=experiment-run:146, feature-engineering:54, build-model:19
cost=$498.31; metrics t0001=.781/1.143 -> t0270=.911/.802
\end{lstlisting}

\end{document}